\documentclass[journal]{IEEEtran}

\ifCLASSINFOpdf
   \usepackage[pdftex]{graphicx}
\else
\fi
\begin{document}
%
\title{Temporal divergence in cropping pattern and its implications on geospatial drought assessment}
%
%
%

\author{C. S. Murthy,M.V.R. Sesha Sai,M. Naresh Kumar,P. S. Roy

\thanks{National Remote Snsing Centre, Hyderabad,
Telangana, 500 037 India e-mail: (murthy\_cs@nrsc.gov.in).}
}

\maketitle

\begin{abstract}
Time series data on cropping pattern at disaggregated level were analysed and its implications on geospatial drought assessment were demonstrated. An index of Cropping Pattern Dissimilarity (CP-DI) between a pair of years, developed in this study, proved that the cropping pattern of a year has a higher degree of similarity with that of recent past years only and tends to be dissimilar with longer time difference. The temporal divergence in cropping pattern has direct implications on geospatial approach of drought assessment, in which, time series NDVI data are compared for drought interpretation. It was found that, seasonal NDVI pro?les of drought year and normal year did not show any anomaly when the cropping patterns were dissimilar and two normal years having dissimilar cropping pattern showed di?erent NDVI profiles. Therefore, it is suggested that such temporal comparisons of NDVI are better restricted to recent past years to achieve more objective interpretation.
\end{abstract}


%
\IEEEpeerreviewmaketitle

\section{Introduction}
%
%
%
%
\IEEEPARstart{F}{rom} the agriculture perspective, drought is a condition in which the amount of water needed for transpiration and direct evaporation exceeds the amount available in the soil. Detection of the incidence and persistence of drought conditions and quantification of its impact on crops still remain as major challenges to the Federal Governments. As a result, development and implementation of efficient, sustainable and economically viable drought management strategies tend to become difficult tasks for the administration.

Currently, agricultural drought conditions are characterized by periodic ground observations of rainfall, aridity and agricultural conditions in terms of cropped area and yield in many countries (Ahmed et al. 2005, Unganai and Bandson 2005, Roy et al. 2006). Rainfall anomalies are derived by comparing the actual rainfall with long-term averages and interpreted for assessing meteorological drought (www.imd.gov.in). The agricultural conditions are monitored through field inspections for making assessments on sowing pattern – time of sowing, extent of sown area, etc., progression of crop growth and crop yield (http://agri.ap.nic.in, http://dmc.kar.nic.in). Thus, near real-time monitoring of rainfall and agricultural situation, provides important source of information for in-season drought assessment, although it encounters with certain obvious limitations like non-spatial nature, inadequate coverage, subjective observations, insufficient datasets, etc. In
India, crop weather watch meetings are held at every state headquarters, on weekly/fortnightly basis during monsoon season to review the agricultural situation in each state/district (www.agricoop.nic.in, http://agri.ap.nic.in). The agro advisory service of India Meteorological Department (IMD) integrates the weather parameters on rainfall, temperature with crop condition and offers suggestions to the farming community in different regions on crop management (http://imdagrimet.org). Thus, the conventional approach of drought assessment is standalone in nature, in the
sense that the decision making on drought prevalence is done based on in-season observations without comparing with any specific reference year or years.

There is another approach for drought assessment being widely adopted in recent years – the geospatial approach – in which the biophysical parameters derived from geospatial images are compared with long-term datasets representing different scenarios like different intensities of drought, normal, better than normal, etc., to assess the anomalies and interpret such anomalies in terms of drought severity. The satellite derived vegetation condition and phenology are proved to be potential indicators for vegetation monitoring and drought assessment (Tucker et al. 1985). The NOAA AVHRR NDVI datasets have been used extensively world over for crop condition monitoring, crop yield assessment and drought detection (Beneditti and Rossini 1993,
Moulin et al. 1998, Peters et al. 2002).

The drought monitor of USA using NOAA-AVHRR data (Brown et al. 2002,
www.cpc.ncep.nooa.gov, www.drought.unl.edu/DM), Global Information and
Early Warning System and Advanced Real Time Environmental Monitoring
Information System of FAO using Meteosat and SPOT–VGT data (Minamiguchi
2005), International Water Management Institute's drought assessment in south west Asia using Modis data (Thenkabail et al. 2004), India's National Agricultural Drought Assessment And Monitoring System (NADAMS) project (Roy et al. 2006, Murthy et al. 2007) are proven examples for operational drought assessment using geospatial information.

Cropping pattern is the prerequisite information for drought assessment, because the impact of drought is largely dependent on the types of crops being cultivated. Crop sown area and cropping pattern are the immediate manifestations of drought situation. The immediate effect of drought is reflected in terms of delay in sowing time or reduction in sown area or changes in cropping pattern. Sowing-related indicators – time of sowing and area sown – are physical in nature and represent specific times in the
season. Cropping pattern – the area under different crops expressed as per cent of total crop area – drives the progression of crop growth, determines the water needs from time to time and critical stages, decides the schedule of operations till harvest, and, hence subjected to cascading effects of drought till the end of season.

Therefore, cropping pattern analysis in terms of its temporal changes –
similarities or divergence, response to drought – complements the in-season drought assessment in many ways. It is a vital input for developing agro-advisory services and for assessing the drought impact. As cropping pattern is the primary input to estimate the seasonal water demand, its temporal changes helps to build vulnerability profile of the area. Marteniz-Casanovas et al. (2005) proposed a method for mapping cropping patterns using time series satellite images and derived year wise crop maps to study cropping pattern variations. Although there are many studies conducted on the cropping pattern variability over time, the results have been linked to socioeconomic aspects such as farm-related activities, livelihoods, shifting of economic opportunities, rural labour migrations, etc. (Walker and James 1990, Bhalla and Gurmail 2001). Keeping in view the importance of temporal cropping pattern information for drought assessment, a detailed analysis of cropping pattern changes and its implications on drought assessment was conducted in this study.

\section{Objectives of the study}
The specific objectives of the study include:
\begin{enumerate}
\item To study the temporal variations of crop sown area and the impact of drought on crop sown area at disaggregated level.
\item To develop and evaluate an index to assess the temporal divergence in cropping pattern at disaggregated level.
\item To identify the years of similar or dissimilar cropping pattern on the basis of index.
\item To discuss the implications of cropping pattern changes on geospatial drought assessment methods which involve comparison of time series datasets.
\end{enumerate}

\section{Study Area}
Mahaboobnagar district, one of the drought prone districts of Andhra Pradesh state, India, is geographically located between 15$^{\circ}$ 55' and 17$^{\circ}$ 20' latitude and 77$^{\circ}$ 15' and 79$^{\circ}$ 15' longitude. The district is generally hot with daily temperatures ranging from 16.9 to 41.58C. The total rainfall in the south west monsoon season is about 600 mm. The district has an area of 1,843,200 ha. Forests occupy 303,000 ha and constitute about 16\% of geographic area. Net sown area is 749,526 ha and forms 40\% of geographic area. Kharif is the main agricultural season in the district and corresponds to June–October/November period. Within the district, there are 64 administrative units called blocks. Blocks form the spatial unit for drought assessment and relief management by the state administration. The study area district with block boundaries inside is shown in Figure 1.
\begin{figure}[!t]
\centering
\includegraphics[width=3in]{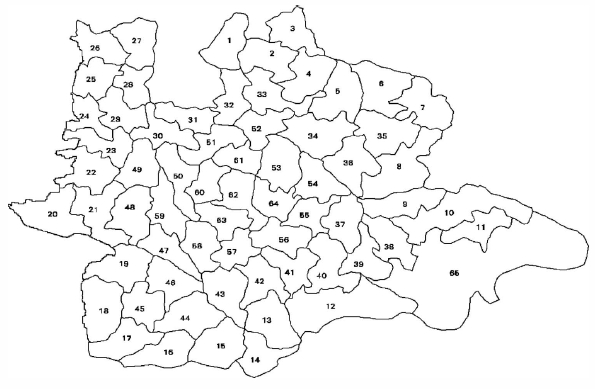}
 \caption{Study area –Mahaboobnagar district with block boundaries inside. (Blocks are the administrative units within district. This district has 64 blocks plus one block with reserved forests.)}
\label{fig1}
\end{figure}

\section{Data used and methodology}
There are three important parts in the analysis – total crop sown area analysis, cropping pattern change analysis and impact of cropping pattern changes on the geospatial drought assessment (Figure 2).

The area under different crops and total crop area at block level during kharif season, for seven years (2000–2006), constitutes the input data base for the analysis. The data were collected from Directorate of Economics and Statistics, Government of Andhra Pradesh, Office of the Chief Planning Officer, Mahaboobnagar district.

Monthly time composite NDVI images of IRS 1C/1D Wide Field Sensor (180 m)
and ResourceSat-1 Advanced Wide Field Sensor (60 m) produced under NADAMS project of National Remote Sensing Centre, Department of Space, Government of India were used to study the geospatial approach of drought assessment. NDVI
images covering the present study area district from the state NDVI images were created for 2006 (drought year), 2005 (normal year) and 2000 (normal year) for further analysis. Seasonal NDVI profiles from June to November were generated at block level for the selected years. Details of the NDVI datasets are available in Murthy et al. (2007).

\begin{figure}[!t]
\centering
\includegraphics[width=3in]{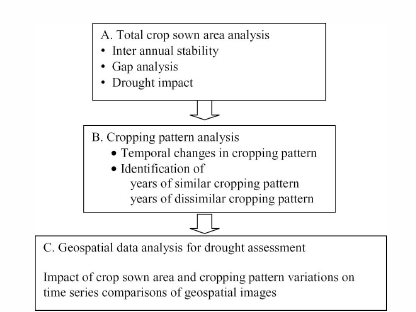}
 \caption{Data exchange mechanisms between the IMS software and interactive, automated work centers}
\label{fig2}
\end{figure}

The cropping pattern in the study area district is dominated by rainfed crops. Only 20\% of cropped area in monsoon season is under irrigation. Rice is the main irrigated crop and it has significant area only in a few blocks located on the southern side of the district. Castor (Pyrgulopsis castor), maize (Zea mays), jowar (Sorghum vulgare), groundnut (Arachis hypogea), sunflower (Helianthus annuus), red gram (Cajanus cajan) and black gram (Craterellus fallax) are the main crops. The cropping pattern of the district is diverse in nature and is dominated by rainfed crops.
\subsection{Total crop sown area analysis}
The inter-annual variability in total crop sown area during the study period was estimated through the coefficient of variation (CV), expressed as standard deviation/mean, in per cent mode. The CV values were calculated at block level. The gap in sown area at block level was calculated as the difference between historic maximum sown area and historic minimum sown area during the period. The normal sown area for each block was calculated as simple average of sown area in normal years. The drought impact on sown area in each block was measured as per cent deviation of actual sown area in drought year from normal sown area.
\subsection{Cropping pattern analysis}
The cropping pattern difference between a pair of years is derived through an index called Cropping Pattern Dissimilarity Index (CP-DI). The index takes the area under different crops expressed as per cent of total crop area of the season for each year as input. The $CP-DI_{jk}$ is computed between the pair of years j and k as under;
\begin{equation}
CP-DI_{jk}=\sum_{i=1}^{n} abs(Y_{ij}-Y_{ik})
\end{equation}
where $Y_{ij}$ per cent area under crop i in year j, $Y_{ik}$ per cent area
under crop i in year k, n = number of crops.

The values of CP-DI range from 0 – representing perfectly similar cropping
pattern to 100 x n – representing mutually exclusive or highest degree of
dissimilarity in cropping pattern.

The crop areas in each year are expressed as per cent of total crop area of the season. Therefore, the CP-DI value can be interpreted as proportion of agricultural area, in which there is a change in the crop types in two years. The CP-DI value of 20 for the years 2006 and 2005 means that in 20\% of agricultural area, the crops cultivated during these two years were different. In 80\% of the area, the crops cultivated were same in these two years. The higher the value of the index, the more is the dissimilarity in the cropping pattern between two years. The transition value of CP-DI between similarity and dissimilarity is taken as 20, in this study. All the CPDI
values above 20 are taken as dissimilar cropping pattern years and $\leq$20 as similar cropping pattern years. The cut-off point of 20 is adopted from the practice of IMD that rainfall deviation of less than or equal to -20\% is considered as meteorological dry period (www.imd.gov.in).

\section{Results and discussion}
\subsection{Inter-annual variability in crop sown area}
The total crop area constitutes the summation of areas under different crops in the season, in any given administrative unit, say, village, block, tehsil or district. The drought situation in the beginning of the season caused by deficit rainfall situation is manifested immediately in the form of either reduction in crop area or delay in sowing time. The time of onset of monsoon rains during June and progression of rainfall up to August determine the extent of sown area and progression of sowing in a given area. In rainfed agriculture, monsoon is a major determinant of crop sowings. Once triggered by rainfall, the sowing pattern involving the selection of suitable crops, time of sowings is determined by other factors like availability of
inputs, the social and economic status of farmers. The assessment of the intensity of drought situation in the first one–two months of the season is done by the approximate estimates on crop sown area.

The temporal changes in the crop sown area in monsoon season during the
period of seven years, was assessed using coefficient of variation (CV). The CV values of different blocks of the study area district are depicted in Figure 3. The CV values of different blocks represented a wide range from >40 to <10\%. The higher the value of CV, the greater the inter-annual variability and hence the more the instability. That means, the extent of the crop sown area during the season shows significant changes from year to year. Histogram of CV shows that the CV was 20–30\% in about 40\% of blocks and it was 30\% and above in about 40\% of blocks. Only in about 15\% of blocks, the CV values were less than 10\%, indicating interannual stability in cropping pattern. Therefore, in majority of the blocks, say, more than 80\% of blocks, the crop area changes were significant, with moderate changes (10–20\% CV) in 44\% of blocks and the higher degree of changes (>20\% CV) in 39\% of blocks. The crop sown areas in different years, in block no. 10 (Achampet) with highest variability (CV = 46\%) ranged from 10,683 ha in a normal year (2000) to 3095 ha in drought year (2003).

\begin{figure}[!t]
\centering
\includegraphics[width=3in]{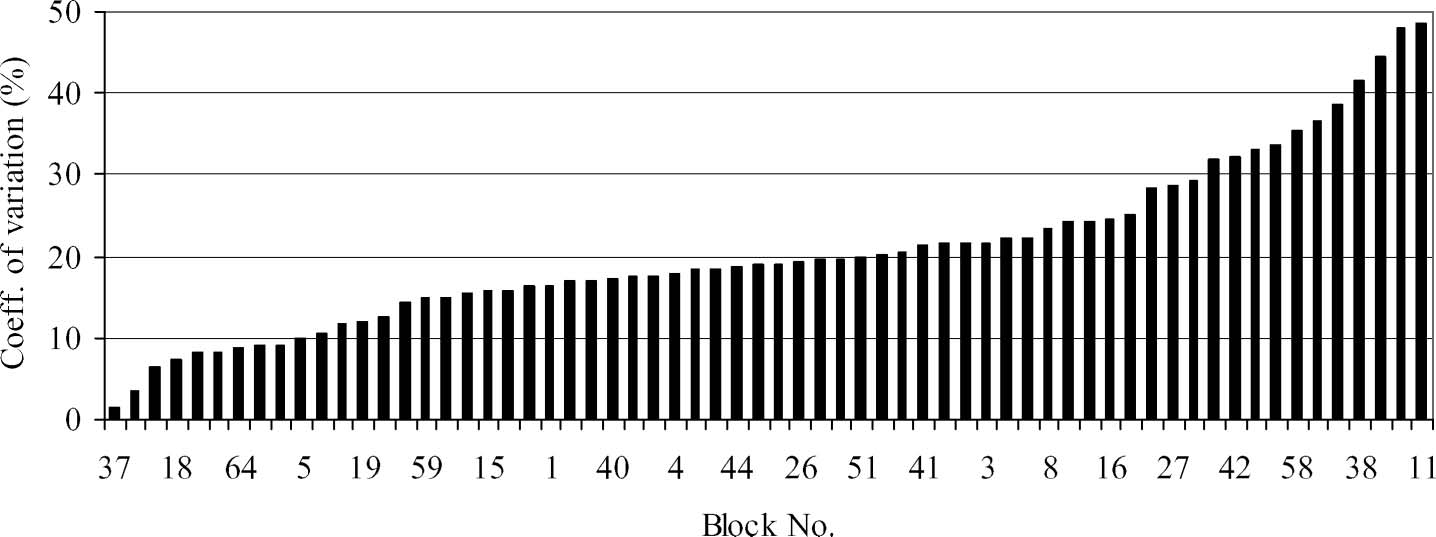}
 \caption{Inter-annual variability (coefficient of variation) in crop sown area.}
\label{fig3}
\end{figure}

\subsection{Gap in total sown area (maximum–minimum)}
The gap between maximum crop sown area and minimum crop sown area (max–
min) indicates the degree of oscillations or sensitivity over time. The maximum and minimum values for different blocks are plotted in Figure 4, which indicates significant crop area fluctuations from year to year in most of the blocks. The histogram of 'max–min' values indicates that the minimum crop sown area is either half or less than half of the maximum crop area in most of the blocks, thus showing potential for very high degree of variability. In most of the blocks, the gap was more than 30\%. The gap of 10–20\% could be seen only in 6\% of blocks, i.e. total crop sown area is stable only in a very few number blocks of study area district.

\section{Impact of drought on crop sown area}

The years 2002 and 2006 received significantly less than normal rainfall during monsoon season and hence the state administration had declared them as drought years. Changes in the extent of crop sown area and time of sowing are the immediate manifestations of rainfall deficiency in the beginning of the season. The response of crop sown area to the droughts of 2002 and 2006 was analysed in terms of per cent reduction in sown area compared to normal. In the majority of the blocks, there was a reduction in sown area during both the years. The histogram of per cent reduction (Figure 5), indicates that in 51\% of blocks in 2006 and 39\% of blocks in 2002, the sown area had reduced significantly from normal. More than 20\% reduction was observed in 10\% of blocks in 2002 and 19\% of blocks in 2006. The drought impact on crop sown area was more pronounced in 2006 than in 2002. In these two years, there were also some blocks with no change in sown area compared to normal or with more than normal sown area. Thus, sown area changes are sigificant even within the district showing different intensities of drought impact.

Thus, the foregoing analysis infers: (a) the total crop sown area in kharif season shows significant variations from year to year in most of the blocks in the study area district, (b) the difference between maximum crop area and minimum crop area is very large in most of the blocks, giving scope for wider fluctuations from year to year, (c) changes in agricultural area from year to year leads to associated changes in current fallow lands and (d) sown area reduction due to drought is very significant and varies among the blocks. A year with maximum area under crops would have minimum area under current fallow lands and vice versa. The process of interchange between agricultural area and current fallow lands is caused by various factors like weather, farmers' preferences, input availability, etc.
\begin{figure}[!t]
\centering
\includegraphics[width=3in]{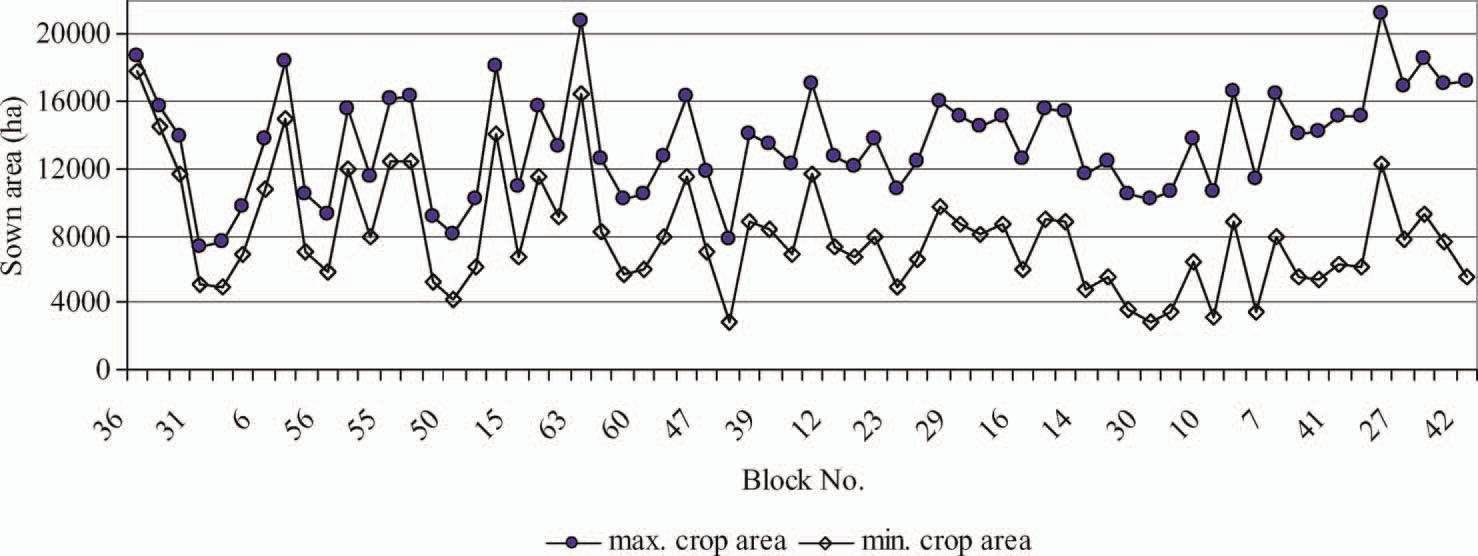}
 \caption{Gap in total crop sown area (maximum–minimum) during 2000–2006.}
\label{fig4}
\end{figure}

\begin{figure}[!t]
\centering
\includegraphics[width=3in]{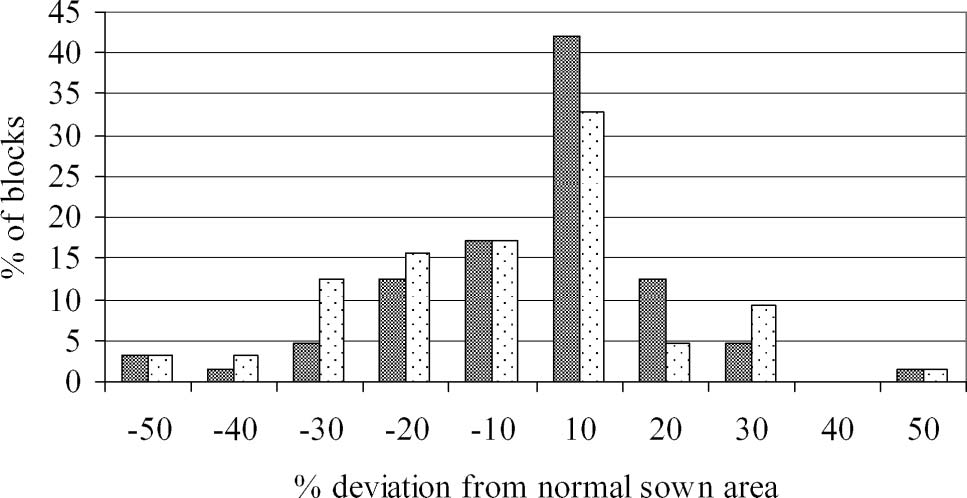}
 \caption{Histogram of sown area changes in drought years}
\label{fig5}
\end{figure}

\subsection{Cropping pattern divergence}
Cropping pattern means the proportion of area under different crops. The total crop area is represented by different crops in the season and the per cent area under different crops in a season signifies the cropping pattern. Cropping pattern change from one year to the other indicates the changes in the proportions of different crops. Some new crops may replace the existing crops. The crops which represent larger proportions are termed as major crops in the area. Cropping pattern of an area is largely determined by resources available such as rainfall, irrigation, soils, farm inputs and ultimately the preferences of farmers. Cropping pattern is an important factor in the process of drought incidence and its persistence. As a part of drought
management strategies farmers prefer to cultivate drought resistant crops like castor, jowar, redgram, etc. in the study area district. Along with the choice of crops, farmers also tend to change the sowing time to escape drought situation. Change in cropping pattern brings changes in crop calendar, total water requirements, chronological occurrence of different crop growth stages, critical stages, etc.

In the study area district, there are two groups of blocks – first group consisting of 11 blocks with <10\% inter-annual variability in crop sown area and 53 number of blocks with large inter-annual variability. In this section, analysis was done to examine the variability in cropping pattern from year to year in these two groups. The total crop area in a given block or district has two components – the area which is actually sown, i.e. total crop sown area and the area which is left unsown, i.e. current fallow lands. The reasons for leaving the land unsown could be unfavourable weather, farmers' decision making, etc. Total sown area and current fallow area keep interchanging from year to year. In all the blocks of the study area
district where the inter-annual change in total sown area is large and there is large gap between potential minimum crop area and potential maximum crop area, this rate of interchanging between crop sown area and the current fallow area is very significant. That means, per cent area under current fallow lands keeps changing significantly from year to year. Therefore, in this group of blocks, dissimilarity in cropping pattern between years was primarily caused by significant inter-annual variability in total sown area and the resultant current fallow lands.

Assessment of inter-annual changes in cropping pattern and evaluation of
divergence or convergence of cropping pattern during the period was done through an index CP-DI as mentioned in previous sections. One block with stable crop area (Telkapally) and one block with unstable crop area (Achampet) were randomly selected for further analysis. The CP-DI matrix for Telkapally block is presented in Table 1.
\begin{table}

\begin{tabular}{c||c||c||c||c||c||c}
  \hline
     Years 	&	 2000 	&	 2001 	&	 2002 	&	 2003 	&	 2004 	&	 2005 	\\	\hline
 2001 	&	 15 	&	 	&	 	&	 	&	 	&	 	\\	\hline
 2002 	&	 16 	&	 15 	&	 	&	 	&	 	&	 	\\	\hline
 2003 	&	 17 	&	 25 	&	 12 	&	 	&	 	&	 	\\	 \hline
 2004 	&	 34 	&	 34 	&	 21 	&	 17 	&	 	&	 	\\	 \hline
 2005 	&	 36 	&	 42 	&	 27 	&	 19 	&	 10 	&	 	 \\	\hline
 2006 	&	 40 	&	 46 	&	 31 	&	 23 	&	 16 	&	 9 	 \\	\hline

\end{tabular}
\caption{CP-DI matrix showing cropping pattern divergence (Telkapally block).}\label{tab1}
\end{table}
A comparison of 2006 cropping pattern with that of the previous years from 2005 to 2000, through CP-DI, shows increasing dissimilarity with history (Table 1). The index value was lower for 2005 and higher for 2001 and 2000 years. In the case of 2005, the dissimilarity was increasing from 2004 to 2000 and in case of 2004, the dissimilarity was increasing from 2003 to 2000. The same trend could be observed for the remaining years also. The cropping pattern dissimilarity was less only with preceding one–two years. As comparison goes to historic years, the cropping pattern dissimilarity increased with larger time difference. For better understanding of the
index, the actual cropping pattern with small index value, i.e. CP-DI = 9 (2006 vs. 2005) and with large index value, i.e. CP-DI = 46 (2006 vs. 2001) are shown in Figures 6 and 7, respectively. With a smaller index value, the proportions of four major crops namely castor, maize, jowar and cotton did not show significant difference between 2006 and 2005. Therefore, the cropping patterns of these two years are similar. On the other hand, with high index value, the proportions of these four crops showed significant difference between 2006 and 2001. The area under maize crop was >50\% of crop area in 2006 compared with <20\% in 2001, the total crop area being the same in both the years. The per cent area under jowar and castor also showed differences between two years. Thus, the cropping patterns of 2006 and
2001 are dissimilar.
\begin{figure}[!t]
\centering
\includegraphics[width=3in]{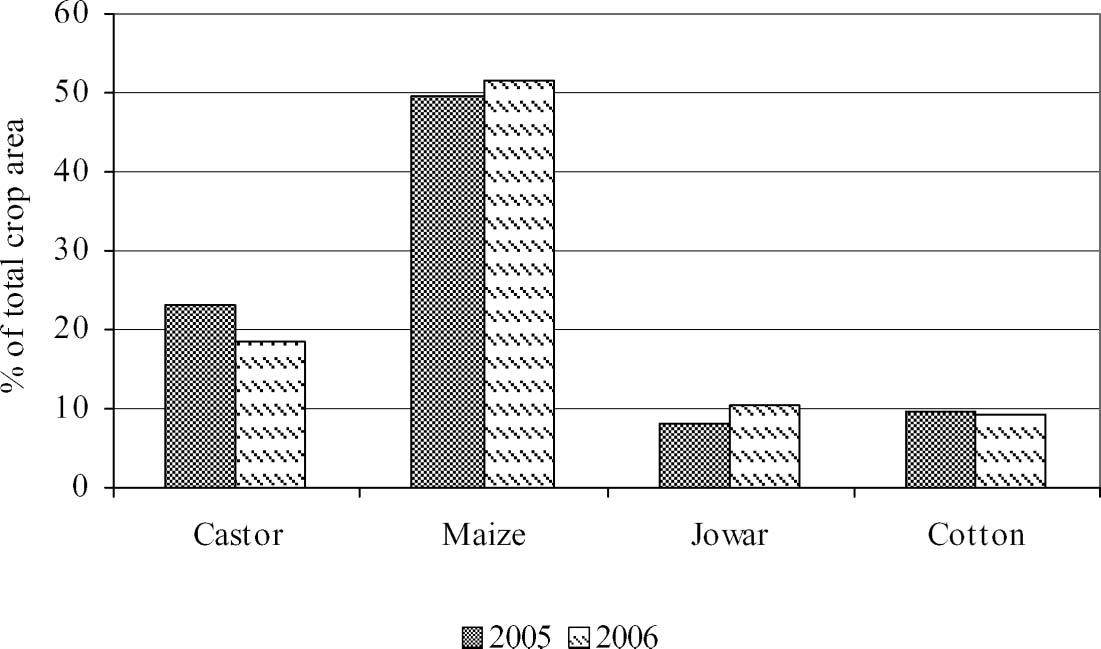}
 \caption{Cropping pattern with small dissimilarity value (CP-DI = 9). (Block no. 37,
Telkapally with least sown area variations from year to year).}
\label{fig8}
\end{figure}

\begin{figure}[!t]
\centering
\includegraphics[width=3in]{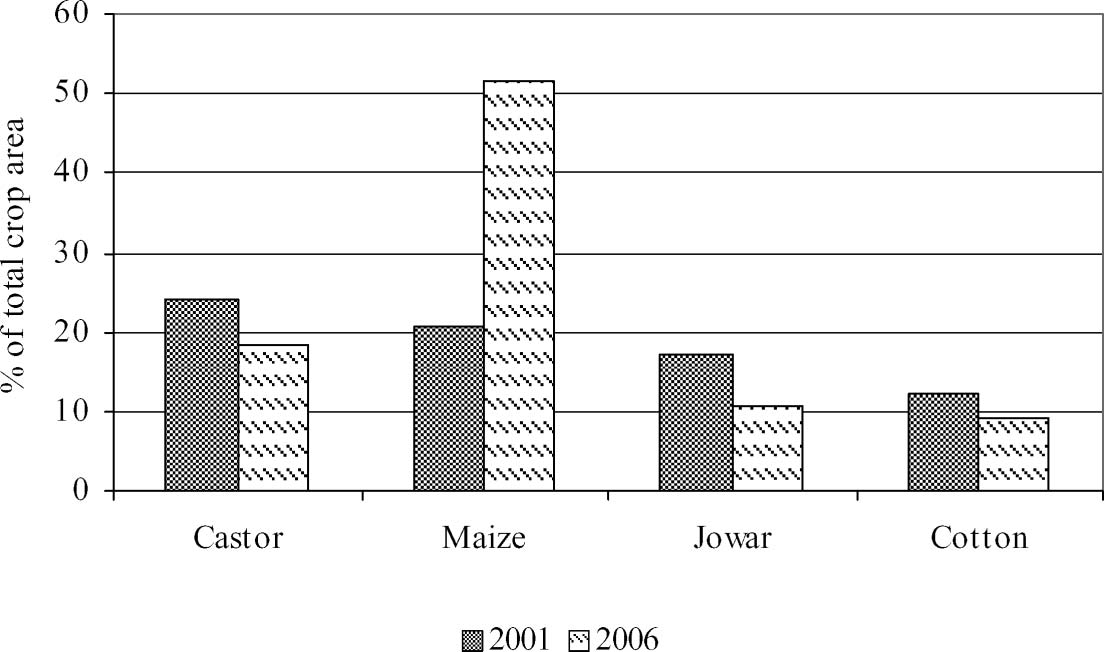}
 \caption{Cropping pattern with large dissimilarity value (CP-DI ¼ 46). (Block no. 37,
Telkapally with least sown area variations from year to year).}
\label{fig9}
\end{figure}

Similar kind of dissimilarity analysis was carried out for another block – Achampeta, which, in contrary to the earlier block has large inter-annual variability in total crop area. The dissimilarity matrix (Table 2), year-to-year comparison of dissimilarity and cropping pattern with low and high index are presented in Figures 8 and 9.
\begin{table}

\begin{tabular}{c||c||c||c||c||c||c}
  \hline
 Years 	&	 2000 	&	 2001 	&	 2002 	&	 2003 	&	 2004 	&	 2005 	 \\	\hline
 2001 	&	 14 	&	 	&	 	&	 	&	 	&	 	\\	\hline
 2002 	&	 71 	&	 65 	&	 	&	 	&	 	&	 	\\	\hline
 2003 	&	 123 	&	 116 	&	 51 	&	 	&	 	&	 	\\	 \hline
 2004 	&	 82 	&	 77 	&	 18 	&	 40 	&	 	&	 	\\	 \hline
 2005 	&	 87 	&	 84 	&	 25 	&	 35 	&	 7 	&	 	\\	 \hline
 2006 	&	 110 	&	 106 	&	 46 	&	 15 	&	 28 	&	 23 	 \\	\hline
\end{tabular}
\caption{CP-DI matrix showing cropping divergence (Achampet block).}\label{tab2}
\end{table}

\begin{figure}[!t]
\centering
\includegraphics[width=3in]{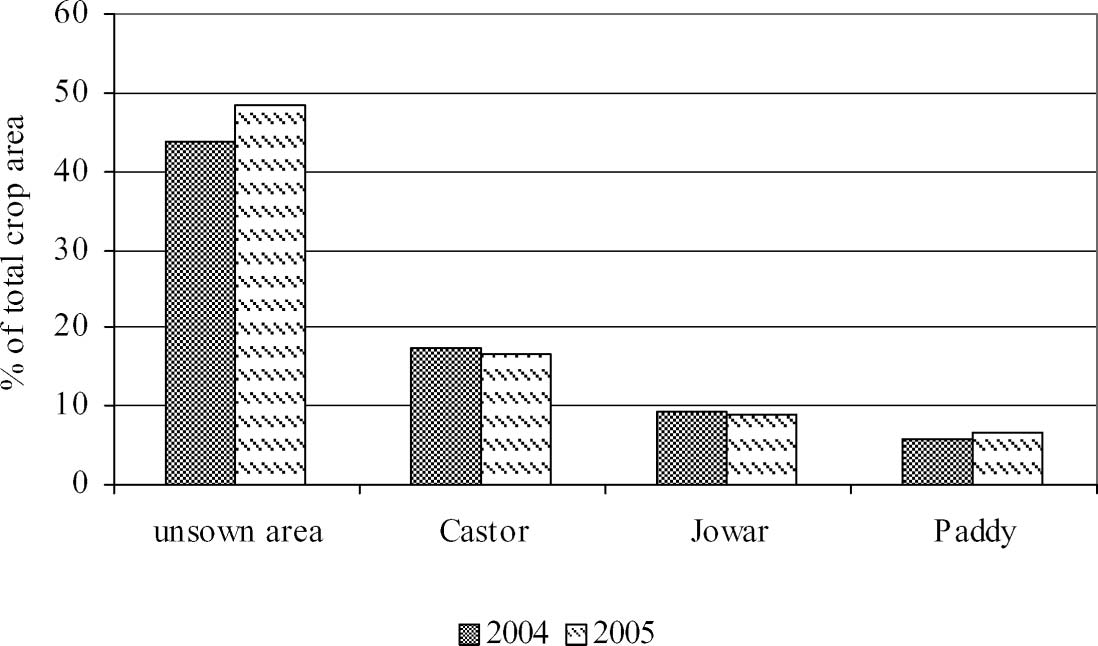}
 \caption{Cropping pattern with less dissimilarity (CP-DI = 7), year 2005 vs. 2004. (Block no. 10, Achampeta with large sown area variations from year to year).}
\label{fig8}
\end{figure}

\begin{figure}[!t]
\centering
\includegraphics[width=3in]{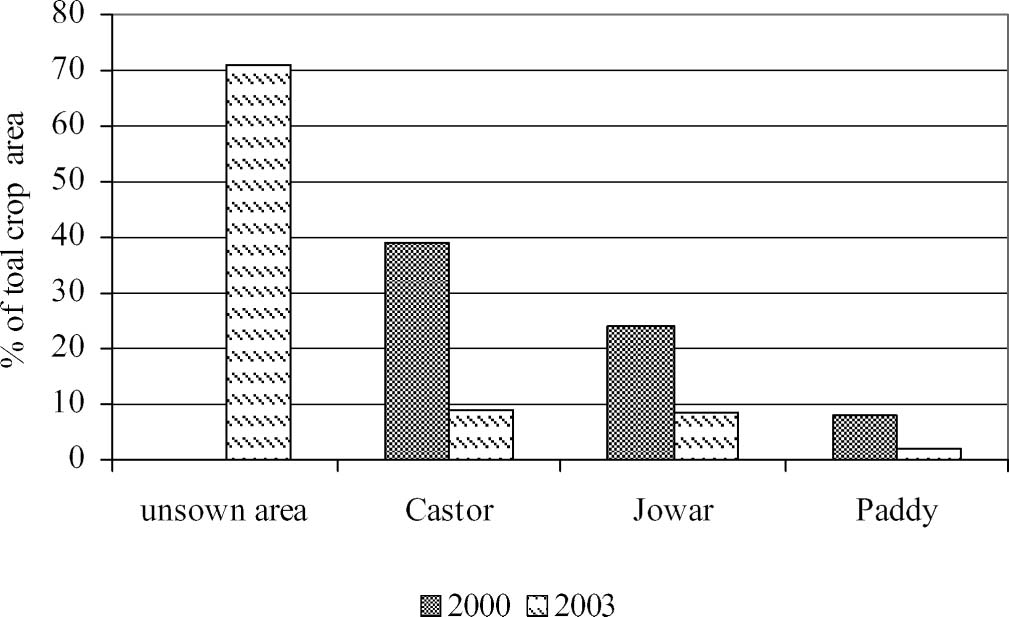}
 \caption{Cropping pattern with high dissimilarity (CP-DI = 123), year 2003 vs. 2000. (Block no. 10, Achampeta with large sown area variations from year to year).}
\label{fig9}
\end{figure}
Because of the significant year-to-year variation in the crop sown area, the
unsown area (current fallow) has become an important component in the cropping pattern of this block. As a result, the crop proportions (\%) for different years are not directly comparable, because of variable total crop sown area from year to year, i.e. the denominator in the calculation procedure. To keep the denominator constant, unsown area is estimated in each year taking the maximum sown area (in the period 2000–2006) as reference. The unsown area or current fallow land was considered as one of the constituents of cropping pattern. In this block also, the dissimilarity tends to increase as the comparisons are extended to more and more previous years. The dissimilarity matrix (Table 2) shows highest dissimilarity in cropping pattern, between 2006 and 2000 and 2006 and 2001 and lower level of dissimilarity with recent past years 2003, 2004 and 2005. Cropping pattern comparisons for each year by pairing with previous years, clearly show that with longer time gap, cropping patterns tend to be more and more dissimilar. The comparison of actual cropping pattern, between 2004 and 2005, with very less index value (CP-DI = 7) as shown in Figure 8 indicates no significant difference in the proportions of different crops and current fallow land between the two years. In contrary to this, the years 2000 and 2003, with very high index values (CP-DI = 123), show significant differences in the
proportions between two years (Figure 9). More than 60\% of area was left unsown (current fallow) in 2003, causing great deal of difference between two years. Also, the area under jowar and castor also differ significantly between the two years. Thus, the years 2000 and 2003 were significantly different with respect to cropping pattern.

Thus, the CP-DI, developed and applied in this study, was sensitive to the
cropping pattern differences between a pair of years. The higher the value of the index, the higher is the dissimilarity and vice versa.

From the foregoing analysis, it is clear that the cropping pattern even in a
localized area like block in this study tends to be dynamic with crop proportions changing rapidly from year to year. The cropping pattern of a given year has higher degree of similarity only with that of recent previous one–two years. During 2006 and 2000, representing a gap of seven years, the change in the cropping pattern was very significant in majority of blocks.

\subsection{Stable crop sown area – diverging cropping pattern}
Out of 64 numbers of blocks in the study area district, only 11 blocks have the stable crop area with coefficient variation less than 10\%, during the period 2000– 2006. The year-to-year variability in the sown area was very minimal in these blocks. There was no significant reduction in the crop sown area in these blocks even in drought years like 2002 and 2006. The cropping pattern changes from year to year were analysed in this section to understand the extent of divergence or convergence over time.

In this group of 11 blocks with stable crop area, three blocks showed unique
patterns of dissimilarity as discussed in the subsequent section. In the remaining eight blocks, four blocks were randomly selected for further analysis and Figure 10 depicts the index values of 2006 versus the rest of the previous years for these four blocks. The index values show that dissimilarity in cropping pattern tends to increase from year 2005 to 2000. The range of dissimilarity values within each year, say, 2006 vs. 2005 or 2006 vs. 2000 is wider. This indicates that different blocks show different levels of dissimilarities with the cropping pattern of previous years. The extent of dissimilarity is not uniform/consistent across time. However, the recent past years tend to have similar cropping pattern with divergence as we move in to extreme past year starting from n73 year in the
current study.
\begin{figure}[!t]
\centering
\includegraphics[width=3in]{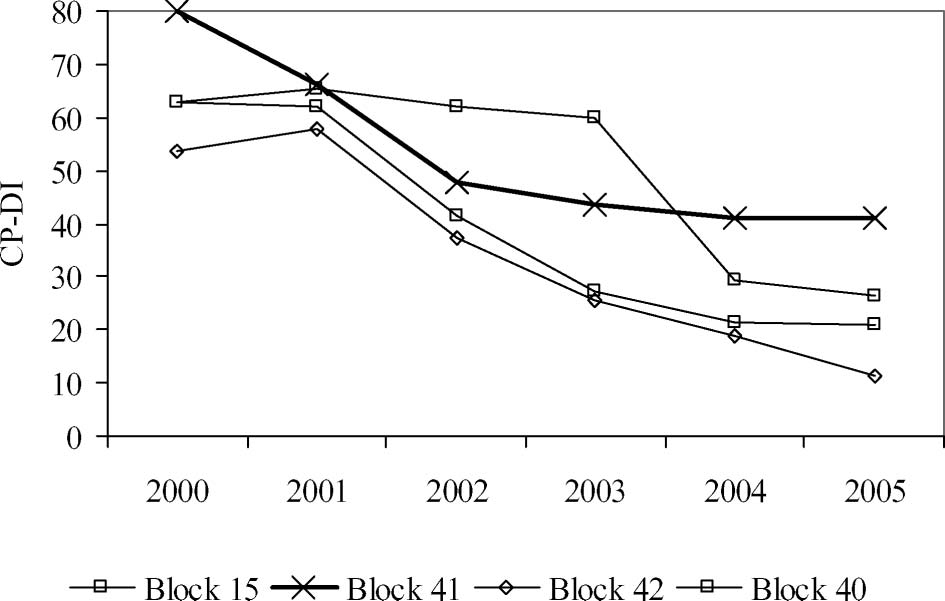}
 \caption{CP-DI-based dissimilarity of 2006 cropping pattern with previous years for selected blocks.}
\label{fig10}
\end{figure}

\subsection{Cropping pattern dissimilarity – some unique patterns}
There are some blocks which showed unique pattern of dissimilarity in cropping pattern. The 2006 cropping pattern of Ghattu block (Figure 11) has a reverse trend, with increasing dissimilarity from 2000 to 2005. The 2006 cropping pattern was very close to that of 2000 and 2001, and tends to be dissimilar with the recent past years, with higher degree of dissimilarity in 2005. The other years of the same block showed a different pattern, more similar to 2005 and dissimilar to 2000.

\begin{figure}[!t]
\centering
\includegraphics[width=3in]{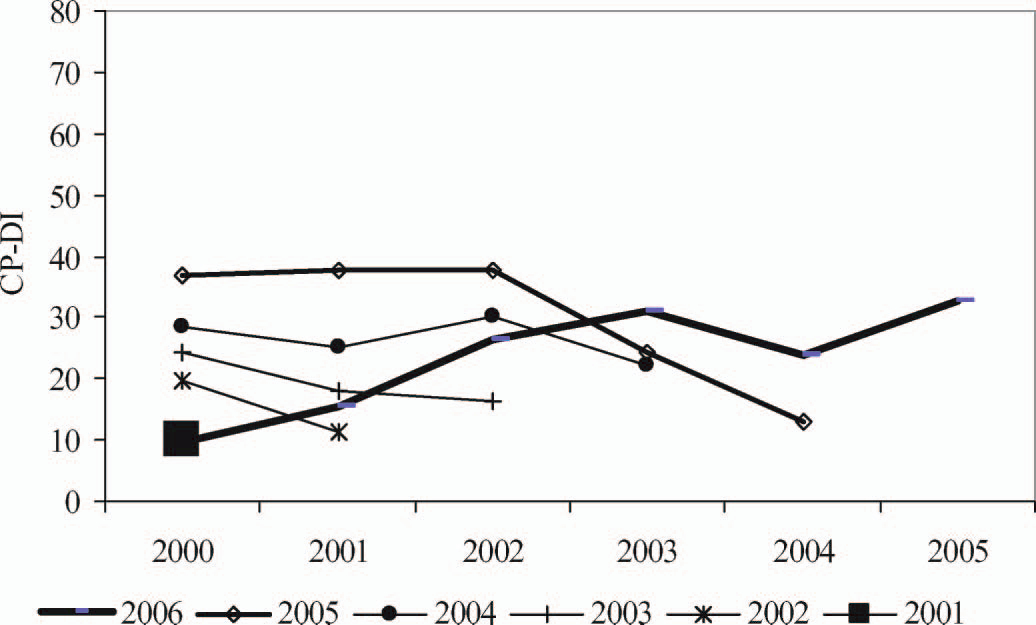}
 \caption{Unique pattern of cropping pattern dissimilarity with previous years – Ghattu block}
\label{fig11}
\end{figure}

In the case of Vidapanakal block, 2006 showed consistently higher level of
dissimilarity with all the past years, indicating that the cropping pattern of 2006 is unique and not comparable with any previous year. In all years, the trend of dissimilarity followed the general pattern as observed in the previous sections, that the cropping pattern was similar to that of recent past years. Again, 2001 cropping pattern was dissimilar to that of 2000, the preceding year.

In the case of Amangal block, the index values were very inconsistent that kept changing from year to year. Although 2006 was close to that of 2005, the index for the remaining years showed higher degree of dissimilarity with wider fluctuations. The year 2005 showed wider fluctuations and not closer to the cropping pattern of any year. The year 2004 has very high level of dissimilarity with all the previous years. The cropping pattern of 2003 was close to that of 2002 and dissimilar to that of 2001 and 2000. Similarly, the years 2002 and 2001 do not exhibit similarity with past years.

\subsection{Identification of the years of similar cropping pattern}
The CP-DI has captured the differences in cropping pattern from year to year, as shown in previous sections. In this section, the pairs of years with similar or dissimilar cropping pattern for the same set of 11 blocks which had stable crop area for the years 2006, 2005 and 2004 were identified on the basis of CP-DI value. The rule applied was that if the CP-DI value for any pair of years is $\leq 20$, the cropping pattern of the two years is similar; otherwise the cropping pattern is dissimilar. Thus, the index values of 2006 in pair with each of previous years from 2005 to 2000 are converted in to two categories namely 'Similar' and 'Dissimilar' and presented in Table 3. Similar exercise for 2005 and 2004 was also done and presented in Tables 4
and 5.
\begin{table}[!t] \tiny
\begin{tabular}{l||l||l||l||l||l||l}
  \hline
  Block &2000 &2001& 2002&2003 &2004 &2005 \\ \hline
 Amangal 	&	 Dissimilar 	&	 Dissimilar 	&	 Similar 	&  Dissimilar 	&	 Dissimilar 	&	 Similar 	\\	\hline
 Talakondapalli 	&	 Dissimilar 	&	 Dissimilar 	&	 Dissimilar 	 &	 Dissimilar 	&	 Similar 	&	 Similar 	\\	\hline
 Gopalpur 	&	 Dissimilar 	&	 Dissimilar 	&	 Dissimilar 	&	 Similar 	&	 Dissimilar 	&	 Dissimilar 	\\	\hline
 Tadoor 	&	 Dissimilar 	&	 Dissimilar 	&	 Dissimilar 	&	 Dissimilar 	&	 Dissimilar 	&	 Dissimilar 	\\	\hline
 Bijinepalli 	&	 Dissimilar 	&	 Dissimilar 	&	 Dissimilar 	&	 Dissimilar 	&	 Similar 	&	 Similar 	\\	\hline
 Veepanagandla 	&	 Dissimilar 	&	 Dissimilar 	&	 Dissimilar 	&	 Dissimilar 	&	 Dissimilar 	&	 Similar 	\\	\hline
 Midgil 	&	 Dissimilar 	&	 Dissimilar 	&	 Dissimilar 	&	 Dissimilar 	&	 Similar 	&	 Similar 	\\	\hline
 Ghattu 	&	 Similar 	&	 Similar 	&	 Dissimilar 	&	 Dissimilar 	&	 Similar 	&	 Dissimilar 	\\	\hline
 Gadwal 	&	 Dissimilar 	&	 Dissimilar 	&	 Dissimilar 	&	 Dissimilar 	&	 Similar 	&	 Similar 	\\	\hline
 Telkapally 	&	 Dissimilar 	&	 Dissimilar 	&	 Dissimilar 	&	 Dissimilar 	&	 Similar 	&	 Similar 	\\	\hline
 Nagarkurnool 	&	 Dissimilar 	&	 Dissimilar 	&	 Dissimilar 	&	 Dissimilar 	&	 Similar 	&	 Similar 	\\	\hline
\end{tabular}
\caption{Similarity\/dissimilarity of 2006 cropping pattern with that of previous years (using
CP-DI values).}\label{tab3}
\end{table}
\begin{table}[!t]\tiny
\begin{tabular}{c||c||c||c||c||c}
  \hline
  Block &2000 &2001& 2002&2003 &2004 \\ \hline
 Amangal 	&	 Dissimilar 	&	 Dissimilar 	&	 Dissimilar 	&	 Dissimilar 	&	 Dissimilar 		\\	\hline
 Talakondapalli 	&	 Dissimilar 	&	 Dissimilar 	&	 Dissimilar 	 &	 Dissimilar 	&	 Similar 		\\	\hline
 Gopalpur 	&	 Dissimilar 	&	 Dissimilar 	&	 Dissimilar 	&	 Similar 	&	 Similar 	\\	\hline
 Tadoor 	&	 Dissimilar 	&	 Dissimilar 	&	 Similar 	&	 Similar 	&	 Similar 	\\	\hline
 Bijinepalli 	&	 Dissimilar 	&	 Dissimilar 	&	 Similar 	&	 Similar 	&	 Similar 		\\	\hline
 Veepanagandla 	&	 Dissimilar 	&	 Dissimilar 	&	 Dissimilar 	&	 Similar 	&	 Similar 		\\	\hline
 Midgil 	&	 Dissimilar 	&	 Dissimilar 	&	 Dissimilar 	&	 Similar 	&	 Similar 		\\	\hline
 Ghattu 	&	 Dissimilar 	&	 Dissimilar 	&	 Dissimilar 	&	 Dissimilar 	&	 Similar 		\\	\hline
 Gadwal 	&	 Dissimilar 	&	 Dissimilar 	&	 Dissimilar 	&	 Dissimilar 	&	 Similar 		\\	\hline
 Telkapally 	&	 Dissimilar 	&	 Dissimilar 	&	 Dissimilar 	&	 Similar 	&	 Similar 		\\	\hline
 Nagarkurnool 	&	 Dissimilar 	&	 Dissimilar 	&	 Similar 	&	 Similar 	&	 Similar 		\\	\hline

\end{tabular}
\caption{Similarity\/dissimilarity of 2005 cropping pattern with that of previous years (using
CP-DI values).}\label{tab4}
\end{table}
\begin{table}[!t]\scriptsize
\begin{tabular}{c||c||c||c||c}
  \hline
  Block &2000 &2001& 2002&2003\\ \hline
 Amangal 	&	 Dissimilar 	&	 Dissimilar 	&	 Dissimilar 	&	 Dissimilar 	\\	\hline
 Talakondapalli 	&	 Dissimilar 	&	 Dissimilar 	&	 Dissimilar 	 &	 Dissimilar 	\\	\hline
 Gopalpur 	&	 Dissimilar 	&	 Dissimilar 	&	 Similar 	&	 Similar 	\\	\hline
 Tadoor 	&	 Dissimilar 	&	 Dissimilar 	&	 Similar 	&	 Similar 	\\	\hline
 Bijinepalli 	&	 Dissimilar 	&	 Dissimilar 	&	 Similar 	&	 Similar 	\\	\hline
 Veepanagandla 	&	 Dissimilar 	&	 Dissimilar 	&	 Dissimilar 	&	 Similar 	\\	\hline
 Midgil 	&	 Dissimilar 	&	 Similar 	&	 Similar 	&	 Similar 	 \\	\hline
 Ghattu 	&	 Dissimilar 	&	 Dissimilar 	&	 Dissimilar 	&	 Dissimilar 	\\	\hline
 Gadwal 	&	 Dissimilar 	&	 Similar 	&	 Similar 	&	 Similar 	 \\	\hline
 Telkapally 	&	 Dissimilar 	&	 Dissimilar 	&	 Dissimilar 	&	 Similar 	\\	\hline
 Nagarkurnool 	&	 Dissimilar 	&	 Dissimilar 	&	 Similar 	&	 Similar 	\\	\hline
\end{tabular}
\caption{Similarity\/dissimilarity of 2004 cropping pattern with that of previous years (using
CP-DI values).}\label{tab5}
\end{table}

The cropping pattern of 2006 was similar to that of either 2005 or 2004 in majority of 11 blocks. A unique situation could be observed in Ghattu block where 2006 is similar to extreme past years (2000 and 2001) and dissimilar with recent past years. In the same way, in Gopalpur block, the cropping pattern of 2006 has similarity only with that of 2003. In Talkondalli block, the year 2006 had no similar years with respect to cropping pattern.

In all the blocks except one, the cropping pattern of 2005 was similar to that of preceding one–two years. In case of Amangal block, 2006 has no similar years (Table 4). In 2004, the cropping pattern was similar to that of preceding one–two years. In three blocks namely, Amangal, Talakondapalli and Ghattu, the year has no previous years with similar cropping pattern (Table 5).
\subsection{Geospatial data for drought assessment}
The geospatial drought assessment is relative in nature – the geospatial vegetation/crop condition images and the resulting index numbers over a geographic area are compared with that of corresponding period of different years representing normal and drought situations. The measured relative deviations are interpreted in terms of drought severity levels. The vegetation condition datasets represent the agricultural situation and hence through comparison of agricultural situations of different years, drought assessment is done (Batista et al. 1997, Unganai and Kogan 1998, Kogan et al. 2003). The term agricultural situation encompasses all the activities related to crops from preparatory cultivation to harvesting. The agricultural situation in a given season is a function of weather parameters, irrigation support and cropping pattern. Although the irrigation support expressed as per cent agricultural area under irrigation, does not change over shorter periods, the cropping pattern being the result of farmers' choices, may tend to change even in relatively shorter periods.

\subsubsection{Implications of cropping pattern differences on geospatial drought assessment}
The impact of cropping pattern differences on time series comparisons of the
seasonal NDVI profiles for anomaly assessment and drought detection is
demonstrated in this section. Comparisons were undertaken under three different situations as detailed below:
\begin{enumerate}
\item  Seasonal NDVI profiles 2006 (drought year) vs. 2005 (normal year)
Figure 12.

\item Case II. Seasonal NDVI profiles 2006 (drought year) vs. 2000 (normal year)Figure 13.

\item Seasonal NDVI profiles of two normal years 2000 vs. 2005 Figure
14.
\end{enumerate}
Significant NDVI anomaly in drought year 2006 from normal year 2005 is
evident from Figure 12. From August onwards when the crops were in active
growing stage, the NDVI values were less than normal. The cropping pattern in the two years was similar with maize (49\% in 2005 and 52\% in 2006), castor (24\% in 2005 and 18\% in 2006), cotton (10\% in 2005 and 9\% in 2006) and jowar (8\% in 2005 and 11\% in 2006). With similar cropping patterns, the differences in NDVI are caused by weather situation. As result, the drought impact is clearly evident from NDVI anomaly.

The NDVI profile of the same drought year 2006 was compared with historic
normal year (2000) as in Figure 13, and the NDVI anomaly was insignificant, thus not reflecting the drought impact. There was a greater degree of cropping pattern differences between the two years, particularly maize was in 52\% of crop area in 2006 and only 25\% in 2000, followed by smaller differences in the areas of castor (27\% in 2000 vs. 18\% in 2006), groundnut (11\% in 2000 vs. 1\% in 2006). The two years under comparison (2005 and 2000) in Figure 14, are normal agricultural years. But their NDVI profiles showed significant differences. The cropping pattern in these two years was dissimilar with maize (25\% in 2000 vs. 49\%in 2005), jowar (14\% in 2000 vs. 8\% in 2006) and groundnut (11\% in 2000 vs. 1\% in 2005).

\begin{figure}[!t]
\centering
\includegraphics[width=3in]{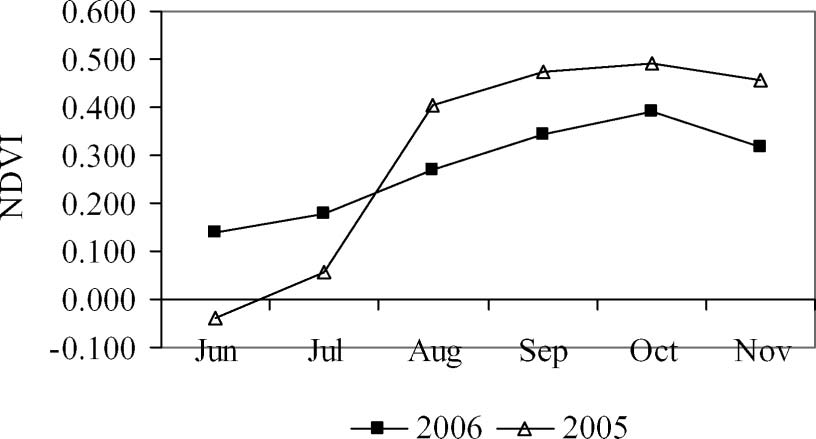}
 \caption{Seasonal NDVI profiles 2006 (drought year) vs. 2005 (normal year).}
\label{fig12}
\end{figure}

\begin{figure}[!t]
\centering
\includegraphics[width=3in]{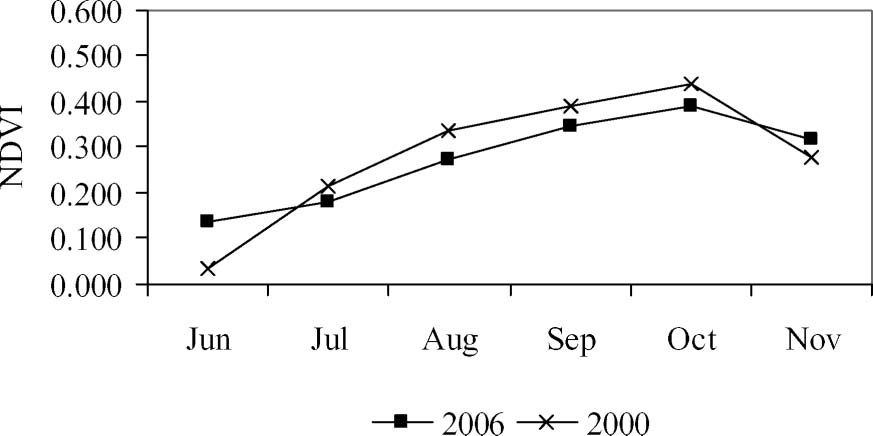}
 \caption{Seasonal NDVI profiles 2006 (drought year) vs. 2000 (normal year).}
\label{fig13}
\end{figure}

\begin{figure}[!t]
\centering
\includegraphics[width=3in]{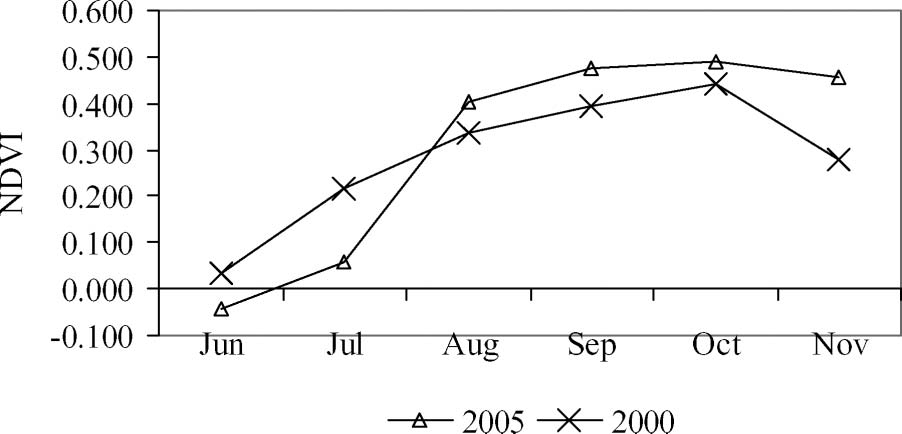}
 \caption{Seasonal NDVI profiles of two normal years 2000 vs. 2005.}
\label{fig14}
\end{figure}

Similar results were obtained when the NDVI profiles were compared with
previous years for other blocks which have insignificant year-to-year sown area differences and the blocks which have significant sown area differences.

Thus, temporal comparison of NDVI profiles for anomaly assessment and
drought detection is subjected to similarity in the cropping pattern between the years of comparison. Two normal years having different cropping pattern showed different seasonal NDVI profiles. Different crops have different spectral response patterns and hence different cropping patterns with combined response tend to produce varying seasonal profiles. Normalization of cropping pattern differences in the aggregated NDVI profiles need parameters on canopy characteristics, spectral response of each crop and hence neither technically feasible nor objective in the operational drought assessment procedures. Instead, it is judicious to select the years having similar
cropping patterns for making such comparisons.

\subsubsection{Implications of inter-annual crop sown area differences}
As the results in previous sections indicated that in the study area district, out of 64 blocks, only in 11 blocks, the total crop sown area remained stable from year to year. In all the other blocks, the year-to-year variability in total crop sown area was very significant. In these blocks, the two constituents of agricultural area – crop sown area and current fallow lands – interchanged from year to year. The time series agricultural area NDVI profiles of a block, used for drought assessment, were influenced by the changing proportions of crop and fallow lands because these two components have contrasting spectral properties. The sensitivity of NDVI profiles
will be enhanced by separating the crop lands and current fallow lands in each season. The crop area layer should be dynamic and season specific, although generation of such layer on real-time basis in the season is hindered by cloud cover problem in the satellite images. Further, delineation of crop areas can be achieved only after complete spectral manifestation of standing crops causing time constraints in operational drought assessment projects.

\section{Conclusions}
From the analysis of crop sown areas and cropping pattern in this study, it is clearly evident that the crop sown areas at disaggregated level, i.e. blocks in a district, changed from year to year even in a short span of seven years. As a result, the interchanges between crop sown area and current fallow lands were very significant in most of the blocks. The huge gap between potential maximum sown area and potential minimum crop area gives scope for large-scale variability in the crop sown area from year to year. There are very less number of blocks with stable crop sown area in the study area district. The index of cropping pattern dissimilarity, CP-DI, developed and applied in this study, is sensitive to the cropping pattern differences
between a pair of years. The higher the index value, the more is the dissimilarity in the cropping pattern between two years. The CP-DI matrix is useful to visualize the cropping pattern variations in a time series data and identify the years having similar cropping patterns. The index finds its application to any dataset to evaluate the temporal cropping pattern changes.

The results of the study have direct implications on geospatial approach of
drought assessment in which, time series datasets on biophysical parameters (e.g. NDVI) are compared for anomaly assessment and drought interpretation. Interannual cropping pattern changes have significant effect on time series NDVI comparisons, as it was observed that seasonal NDVI profiles of drought year and normal year did not show any anomaly when the cropping patterns are dissimilar. Further, two normal years having dissimilar cropping patterns showed different NDVI profiles.

Therefore, selection of years with similar cropping patterns is very important in the time series datasets, to make objective temporal comparisons of bio-physical parameters vis-a`-vis anomalies. This is particularly more relevant to the geospatial approach of drought assessment where long-term datasets are subjected to quantitative comparisons. Results of the current study suggest that such temporal comparisons of vegetation condition be restricted to recent past years, because cropping pattern changes are significant with longer time gap. If cropping pattern remains same between two years, the changes in the crop condition between the two years are attributed to weather changes; otherwise, the change is contributed by both
cropping pattern changes and weather changes.
\%appendices
%
%
%
%
%

\ifCLASSOPTIONcaptionsoff
  \newpage
\fi


\begin{thebibliography}{1}

\bibitem{1}
 Ahmed, U.A., Anwar Iqbal, and Abdul, M.C., 2005. Agricultural drought in bangladesh. Monitoring and predicting agricultural drought – a global study. New Delhi, India: Oxford University Press.
\bibitem{2}
 Batista, T.T., Shimabukuro, Y.E., and Lawrence, W.T., 1997. The long term monitoring of vegetation cover in the Amazonian region of northern Brazil using NOAA-AVHRR data. International Journal of Remote Sensing, 18, 3195–3210.
\bibitem{3}
Benedetti, R. and Rossini, P., 1993. On the use of NDVI profiles as a tool for agricultural
statistics – the case study of wheat yield estimate and forecast in Emilia Romanga. Remote
Sensing of Environment, 45, 311–326
\bibitem{4}
Bhalla, G.S. and Gurmail, S., 2001. Indian agriculture; four decades of development. New Delhi: Sage Publications
\bibitem{5}
 Brown, F.J., et al., 2002. A prototype drought monitoring system integrating climate and
satellite data. In: Proceedings of the Pecora L5/land satellite information 1V/ISPRS
commission I/FIEOS, Colarado, USA.
\bibitem{6}
 Kogan, F.N., et al., 2003. AVHRR based spectral vegetation index for quantitative assessment
of vegetation state and productivity: calibration and validation. Photogrammetric
Engineering and Remote Sensing, 69, 899–906
\bibitem{7}
 Martinez-Casasnovas, A.J., Martin-Montero, A., and Casterad, A.M., 2005. Mapping multiyear
cropping patterns in small irrigation districts from time series analysis of Landsat TM
images. European Journal of Agronomy, 23, 159–169.
\bibitem{8}
Minamiguchi, N., 2005. The application of geospatial and disaster information for food
insecurity and agricultural drought monitoring and assessment by the FAO GIEWS and
Asia FIVIMS. In: Proceedings of the workshop on reducing food insecurity associated with
natural disasters in Asia and the Pacific, Bangkok, Thailand.
\bibitem{9}
 Moulin, S.A., Bondeau, A., and Delecolle, R., 1998. Combining agricultural crop models and
satellite observations: from field to regional scales. International Journal of Remote
Sensing, 19, 1021–1036.
\bibitem{10}
Murthy, C.S., et al., 2007. Agricultural drought assessment at disaggregated level using
AWiFS/WiFS data of Indian remote sensing satellites. Geocarto International, 22, 127–140.
\bibitem{11}
 Peters, J.A., et al., 2002. Drought monitoring with NDVI-based standardized vegetation
index. Photogrammetric Engineering and Remote Sensing, 68, 71–75.
\bibitem{12}
 Roy, P.S., et al., 2006. Geoinformatics for drought assessment. In: J.S. Samra, S. Gurbachan,
and J.C. Dagar, eds. Drought management strategies in India. New Delhi, India: ICAR,
23–60.
\bibitem{13}
 Thenkabail, P.S., Gamage, M.S.D.N., and Smakhtin, V.U., 2004. The use of remote sensing
data for drought assessment and monitoring in Southwest Asia. Srilanka: International
Water Management Institute (IWMI). Research Report No. 85. Available from:
www.iwmi.org [Accessed 15 Feb 2008]
\bibitem{14}
Tucker, C.J., Townshend, J.R.G., and Goff, T.E., 1985. African land cover classification using
satellite data. Science, 227, 369–375.
\bibitem{15}
 Unganai, L.S. and Kogan, F.N., 1998. Drought monitoring and corn yield estimation in
Southern Africa from AVHRR data. Remote Sensing of Environment, 63, 219–232.
\bibitem{16}
Unganai, S.L. and Bandson, T., 2005. Monitoring agricultural drought in South Africa.
Monitoring and predicting agricultural drought – a global study. Oxford and London, UK:
Oxford University Press, 266–275.
\bibitem{17}
Walker, T. and James, R., 1990. Village and household economies in India's semi arid tropics.
Maryland, USA: The Johns Hopkins University Press.



\end{thebibliography}
\end{document}